\documentclass[fleqn,usenatbib]{mnras}

\usepackage[T1]{fontenc}
\usepackage{ae,aecompl}
\usepackage{amsmath}
\usepackage{amssymb}
\usepackage{graphicx}
\usepackage{lastpage}
\usepackage{enumerate}
\usepackage{mathrsfs}
\usepackage{braket}
\usepackage[version=4]{mhchem}
\usepackage{txfonts}

\newcommand\mati{\begin{matrix}}
\newcommand\matf{\end{matrix}}
\newcommand\bmati{\begin{bmatrix}}
\newcommand\bmatf{\end{bmatrix}}
\newcommand\pmati{\begin{pmatrix}}
\newcommand\pmatf{\end{pmatrix}}

\newcommand{\pmat}[1]{\pmati #1 \pmatf}

\newcommand{\abs}[1]{\left\lvert#1\right\rvert}
\newcommand{\pushright}[1]{\ifmeasuring@#1\else\omit\hfill$\displaystyle#1$\fi\ignorespaces}
\newcommand{\pushleft}[1]{\ifmeasuring@#1\else\omit$\displaystyle#1$\hfill\fi\ignorespaces}
\newcommand{\quotes}[1]{``#1''}
\renewcommand{\vec}{\overset{\rightharpoonup}} 

\author[
    J. Forer,
    D. Hvizdo\v{s},
    M. Ayouz,
    C. H. Greene,
    V. Kokoouline
]{
    Joshua Forer$^{1,2}$
    \thanks{Email: \href{mailto:j.forer@posteo.net}{j.forer@posteo.net}},
    D\'avid Hvizdo\v{s}$^3$,
    Mehdi Ayouz$^4$,
    Chris H. Greene$^3$,
    Viatcheslav Kokoouline$^1$
    \\
    \mbox{$^1$Department of Physics, University of Central Florida, 32816, Florida, USA }
    \\
    \mbox{$^2$Institut des Sciences Moléculaires, Université de Bordeaux, CNRS UMR 5255, 33405, Talence Cedex, France}
    \\
    $^3$Department of Physics and Astronomy and
    Purdue Quantum Science and Engineering Institute, Purdue University, West Lafayette, Indiana 47907, USA
    \\
    $^4$Université Paris-Saclay, CentraleSupélec, Laboratoire de Génie des Procédés et Matériaux,
    91190, Gif-sur-Yvette, France.
}

\title[
    Rate coefficients for \ce{e- + CH+}: DR and RVE
]{
    Kinetic rate coefficients for electron-driven collisions with \ce{CH+}: dissociative recombination and rovibronic excitation.
}

\date{\today}

\pubyear{2023}

\begin{document}
\label{firstpage}
\pagerange{\pageref{firstpage}--\pageref{lastpage}}
\maketitle

\begin{abstract}
Cross sections and rate coefficients for rovibronic excitation of the CH$^+$ ion by electron impact and dissociative recombination of CH$^+$ with electrons are evaluated using a theoretical
approach combining an R-matrix method and molecular quantum defect theory. The method has been developed and tested, comparing the theoretical results with the data from the recent Cryogenic Storage Ring experiment. The obtained cross sections and rate coefficients evaluated for temperatures from 1~K to 10,000~K could be used for plasma modeling in interpretation of astrophysical observations and also in technological applications where molecular hydrocarbon plasma is present.
\end{abstract}

\begin{keywords}
    molecular processes --
    plasmas --
    scattering --
    astrochemistry --
    ISM: clouds
\end{keywords}
\maketitle


\section{Introduction}

Hydrides and their ions are often used to trace various characteristics of the interstellar medium (ISM).
Understanding their formation and destruction mechanisms is therefore necessary in using them as accurate tracers.
The \ce{CH+} ion, discovered in the interstellar medium by \cite{douglas1941note} and \cite{adams1941some}, was originally thought to be formed primarily through the reaction
\begin{equation}
    \ce{C+ + H2 -> CH+ + H}.
    \label{eqn:CH+Formation}
\end{equation}
The reaction  is endothermic, requiring 0.398~eV ($\sim$4620~K) to proceed \citep{hierl1997rate}, but typical kinetic temperatures of diffuse clouds are roughly between 40 and 130~K~ \citep{shull2021far}.
Hence, the reaction cannot explain the observed abundance of \ce{CH+} in diffuse clouds \citep{godard2013complete}.
Observations of CH$^+$ in diffuse clouds have motivated many theoretical and experimental studies on the structure and reactivity of the ion. The structure of the ion has been well established for several decades, while there is still need for cross sections for processes involving collisions of CH$^+$ with electrons. In particular, the knowledge of cross sections for excitation of the ion by electron impact is important for interpretations of astrophysical observations. For example, rotational excitation of CH$^+$ by electron impact was found to be the dominant process producing CH$^+$ in the ISM \citet{godard2013complete}.

Besides astronomical applications, cross sections and rate coefficients for \ce{e- + CH+} collisions are important for the interpretation and modelling of hydrocarbon plasma behavior. Other than rotational excitation, mentioned above, one needs the data on vibrational and electronic excitation and dissociative recombination (DR). CH$^+$ is also a suitable candidate for benchmark theoretical studies of such processes.
On one hand, this is because there are several experimental measurements \citep{amitay1996dissociative,paul2022experimental,kalosi2022laser} available -- of DR in particular.
On the other hand, the ion has a relatively complex electronic structure, such that theoretical methods could be tested on this system and, if successful, be applied to other similar problems.

Our recent theoretical study \citep{forer2023unified} has demonstrated that theory can now accurately describe the DR process in CH$^+$.
The theoretical approach developed in that study (originated from several previous works \citep{hamilton2002competition,kokoouline2003unified,vcurik2020dissociative}) with small modifications can also be used to obtain cross sections for rotational, vibrational, and even electronic excitation of the CH$^+$ ion by electron impact.
Because the theoretical method was validated comparing the DR results with the experimental data, it is expected to provide reliable data on the electron-impact excitation processes as well.
This study is devoted to the theoretical evaluation of cross sections and rate coefficients for rovibronic excitation of CH$^+$ ion by electron impact.
The only available  experimental result on excitation of CH$^+$ is from a recent study in the Cryogenic Storage Ring (CSR) \citep{kalosi2022laser}, where the rate coefficient for rotational excitation from the ground rovibrational level $v=0,j=0$ of CH$^+$ to the first excited $v=0,j=1$ was measured.
However, there have been several theoretical studies on rotational excitation using the Coulomb-Born approximation \citep{chu1974rotational,dickinson1977rotational,lim1999electron} and a semi-classical method \citep{flower1979electron}.
More recently, vibronic excitation of CH$^+$ was studied by \citet{jiang2019cross} using a combination of an R-matrix approach \citep{carr2012ukrmol} and molecular quantum defect theory \citep{seaton1983quantum,aymar1996multichannel}.

The present study is based on a fully quantum description of \ce{e- + CH+} collisions and considers rotational and vibrational degrees of freedom of the target ion, as well as its electronic structure, including the three lowest electronic states and corresponding Rydberg resonances appearing in the \ce{CH+ + e-} collisional spectrum.
The article is organized in the following way:
section \ref{sec:theory} gives an overview of the present theoretical approach and the differences between its application for DR and rovibronic excitation,
section \ref{sec:DRRates} presents our DR rate coefficients,
section \ref{sec:coulomb} describes the Coulomb-Born approximation and its application to the present rotational excitation calculations,
section \ref{sec:RVERates} is devoted to the discussion of obtained results on rovibronic excitation, and
section \ref{sec:conclusion} concludes our findings.
\section{Theoretical Approach}
\label{sec:theory}

The method described in this section combines fixed-nuclei electron-scattering calculations with the R-matrix method, rovibrational frame transformation, and multichannel quantum defect theory (MQDT).
Only the main elements of the approach will be presented here.
We use the same method to calculate scattering matrices as in our previous study of the dissociative recombination (DR) of \ce{CH+} \citep{forer2023unified}.
The theory and computational details have only two main differences: the formula for obtaining the rovibronic excitation (RVE) cross sections (instead of DR cross sections) and the vibrational Hamiltonian used in the vibrational frame transformation of the electronic S-matrix.

We perform electron-scattering calculations with the R-Matrix method implemented in UKRMol \citep{carr2012ukrmol,tennyson2010electron}, accessed via the Quantemol-N interface \citep{tennyson2007quantemol}.
The K-matrices for the \ce{e- + CH+} system are represented in a basis of {\it electronic} scattering channels, indexed by $n$, $l$, and $\lambda$.
The quantum numbers $l$ and $\lambda$ correspond to the magnitude of the orbital angular momentum of the incident electron and its projection on the molecular axis, respectively.
The electronic states of the target, $\ce{CH+}$, are indexed by $n$.
K-matrices are obtained for several values of $R$, the internuclear distance of \ce{CH+}.
We then transform the K-matrices into the S-matrix, via the intermediary matrix of phase shifts, $\underline{\delta}(R)$.
The matrices are formally related by
\begin{equation}
  \underline{K}(R) = \tan\underline{\delta}(R),
  \quad
  \underline{S}(R)  = (I+iK)(I-iK)^{-1} = e^{\underline{\delta}(R)}.
  \label{eqn:K_S_delta}
\end{equation}
The reasons for this transformation are twofold: the S-matrix is a smooth function of $R$, which is necessary to have a more accurate vibrational integral in the frame transformation, and the S-matrix is used in the formula to compute the cross sections.

After the electronic S-matrices are obtained for each value of $R$, we proceed with the vibrational frame transformation.
The first difference in this approach with our study of DR \citep{forer2023unified} is that the vibrational Hamiltonian is Hermitian.
To study DR, we used a complex absorbing potential to represent absorption by discretizing the continuum.
Here, we only need to obtain vibrational wave functions for bound states.
The vibrational frame transformation proceeds as
\begin{equation}
  S^\Lambda_{n'v'l'\lambda',nvl\lambda} = \int dR\; \phi_{n'v'}(R)\; S^\Lambda_{n'l'\lambda',nl\lambda}\; \phi_{nv}(R),
  \label{eqn:VFT}
\end{equation}
where $v$ indexes a vibrational level.
The superscript $\Lambda$ indicates that the S-matrices are block diagonal with respect to $\Lambda$, the projection of the total angular momentum on the molecular axis.

The Hermitian Hamiltonian implies, of course, real eigenvalues.
The channel energies in the case of DR are complex, with a nonzero imaginary part for continuum states.
Here, all channel energies and vibrational wave functions $\phi_{nv}(R)$ are real-valued.
The S-matrix on the left-hand-side of (\ref{eqn:VFT}) is the {\it vibronic} S-matrix and, unlike in the case of DR, unitarity is defined with the usual spectral norm, i.e., $S^JS^{J\dagger} = I$.
Following the vibrational frame transformation, we perform the rotational frame transformation on the vibronic S-matrix to obtain the {\it rovibronic} S-matrix, i.e.,
\begin{equation}
  S^J_{n'v'j'\mu'l',nvj{\mu}l} = \sum\limits_\Lambda \sum\limits_{\lambda\lambda'} (-1)^{l'+l+\lambda'+\lambda} C^{j'\mu'}_{l'-\lambda',J\Lambda} S^\Lambda_{n'v'l'\lambda',nvl\lambda} C^{j\mu}_{l-\lambda,J\Lambda},
  \label{eqn:RFT}
\end{equation}
where the total angular momentum of the ion-electron system is $\vec{J} = \vec{j} + \vec{l}$, $\vec{j}$ is the total angular momentum of the ion, and $\mu$ is the projection of $j$ on the molecular axis.
The S-matrix, now expressed in a basis of rovibronic channels, block diagonal over $J$.

For each scattering energy, we then partition $S^J$ into blocks corresponding to open ($o$) and closed ($c$) channels and construct the diagonal matrix $\underline{\beta}$ for closed channels,
\begin{equation}
  \underline{S}^J = \pmat{ \underline{S}_{oo} & \underline{S}_{oc} \\ \underline{S}_{co} & \underline{S}_{cc} },
  \quad
  \beta_{i'i}(E_\text{tot}) = \frac{\pi}{\sqrt{2(E_i-E_\text{tot})}} \delta_{i'i},
  \label{eqn:S_partition}
\end{equation}
where $E_i$ is the energy of the $i^\text{th}$ channel and is real, and $E_\text{tot}$ is the total energy of the ion-electron system.
We proceed with the closed-channel elimination procedure, borrowed from MQDT, to reduce the S-matrix to only open channels.
\begin{equation}
  \underline{S}^{J,phys} (E_\text{tot}) = \underline{S}_{oo} - \underline{S}_{oc} \left( \underline{S}_{cc} - e^{-2i\underline{\beta}} \right) \underline{S}_{co}.
  \label{eqn:S_phys}
\end{equation}
The physical S-matrix, $S^{J,phys}$, is then used to calculate the total RVE cross section from some initial channel $\ket{nvj\mu}$ to some final channel $\ket{n'v'j'\mu'}$,
\begin{equation}
  \sigma_{n'v'j'\mu' \leftarrow nvj\mu} (E_{el}) = \frac{\pi}{2m_eE_{el}} \sum\limits_J \frac{2J+1}{2j+1}  \sum\limits_{ll'} \abs{S^{J,phys}_{n'v'j'\mu'l',nvj\mu l}}^2,
  \label{eqn:XS_rot}
\end{equation}
where $m_e$ is the mass of an electron and $E_{el}$ is the incident electron energy.
It is also possible to calculate vibronic excitation (VE) cross sections, i.e., not including the rotational structure, by simply skipping the rotational frame transformation (\ref{eqn:RFT}).
The closed-channel elimination procedure remains identical, except that the S-matrices are block diagonal over $\Lambda$ and not $J$.
The total VE cross section from some initial channel $\ket{nv}$ to some final channel $\ket{n'v'}$ is then
\begin{equation}
  \sigma_{n'v' \leftarrow nv} (E_{el}) = \frac{\pi}{2m_eE_{el}} \sum\limits_{ll'}\sum\limits_{\lambda\lambda'} \abs{S^{\Lambda,phys}_{n'v'l'\lambda',nvl\lambda}}^2.
  \label{eqn:XS_norot}
\end{equation}

The cross sections obtained from (\ref{eqn:XS_rot}) and (\ref{eqn:XS_norot}) only describe a single scattering event.
To better describe conditions in the ISM, kinetic rate coefficients are needed, which rely on the above cross sections.
State-selected kinetic rate coefficients, for DR, RVE, or VE, are obtained following
\begin{equation}
  \label{eqn:kinetic_rate_select}
  \alpha_k(T) = \frac{
    \int\limits_0^\infty \sigma(E_{el}) \sqrt{2E_{el}/m_e} \sqrt{E_{el}} e^{-E_{el}/kT} dE_{el}
  }{\int\limits_0^\infty \sqrt{E_{el}} e^{-E_{el}/kT} dE_{el}},
\end{equation}
where $k$ is the Boltzmann factor and $\sigma$ is a cross section.
In practice, these integrals are carried out numerically.
Additionally, one can average the state-selected rate coefficients obtained from (\ref{eqn:kinetic_rate_select}) by
\begin{equation}
  \overline{\alpha}_k(T) = \frac{
    \sum\limits_{i} \alpha_k^i(T) (2j_i+1) e^{-E_{i}/kT}
  }{
    \sum\limits_{i} (2j_i+1) e^{-E_{i}/kT}
  },
  \label{eqn:kinetic_rate_avg}
\end{equation}
where $i$ indexes starting channels where the total angular momentum quantum number of the ion is $j_i$.
The rate coefficient (DR or (R)VE), starting from some channel indexed by $i$ and obtained with (\ref{eqn:kinetic_rate_select}), is given by $\alpha_k^i(T)$.
If $\alpha_k^i(T)$ is a rate coefficient without rotational resolution, $j_i$ can be taken to be zero in (\ref{eqn:kinetic_rate_avg}).

The precision of theoretical cross sections are only limited by the numerical precision of the calculations.
Experimental measurements have much larger uncertainties, so comparisons are often made by convolving theoretical results with experimental parameters.
The convolution function differs for every experimental setup, but Gaussian functions are fairly common:
\begin{align}
    \overset{\sim}{\sigma}(E) &= \frac{
      \int dE \sigma(E_{el}) e^{-\left(E_{el}-E\right)^2 / (2\gamma^2) }
    }{
      \int dE e^{-\left(E_{el}-E\right)^2 / (2\gamma^2)}
    }
    \label{eqn:convolution_gauss_numerical}
    \\
    \overset{\sim}{\sigma}(E)
    &=
    \frac{1}{\gamma\sqrt{2\pi}} \int dE \sigma(E_{el}) e^{-\left(E_{el}-E\right)^2/ (2\gamma^2) }.
    \label{eqn:convolution_gauss_analytic}
\end{align}
The parameter $\gamma$ is the convolution width (in the same energy units as the electron energy grid).
The prefactor in (\ref{eqn:convolution_gauss_analytic}) is the analytic expression of the denominator in (\ref{eqn:convolution_gauss_numerical}).
Because calculations are performed numerically on predetermined grids of scattering energies, (\ref{eqn:convolution_gauss_analytic}) and (\ref{eqn:convolution_gauss_numerical}) may give different results near the endpoints.

\section{Rate Coefficients for Dissociative Recombination}
\label{sec:DRRates}

\begin{figure}
  \centering
  \includegraphics[width=.48\textwidth]{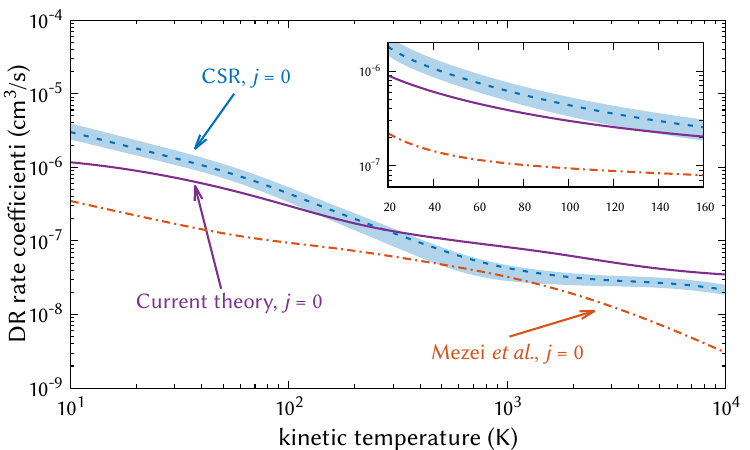}
  \caption{
    State-selected DR cross sections from the ground vibronic state of \ce{CH+}.
    The solid line was obtained with (\ref{eqn:kinetic_rate_select}) using the theoretical DR cross sections obtained with the current method \protect\citep{forer2023unified}.
    The dashed line with an error-curve represents recent experimental data \protect\citep{paul2022experimental}.
    The dot-dashed line represents previous calculations \protect\citep{mezei2019dissociative}.
  }
  \label{fig:DR}
\end{figure}

Fig. \ref{fig:DR} compares state-selected kinetic DR rate coefficients obtained with the present method \citep{forer2023unified}.
DR and RVE are related in the sense that they are competing processes.
During a collision between an ion and an electron, if the initial channel is not the only open channel, both processes may take place.
This is why we take the DR probability to be the probability that no RVE occurs.
Therefore, DR results exhibiting a certain level of agreement with experimental results would suggest that RVE is described, overall,  with similar accuracy.
The present method produces much more accurate kinetic DR rate coefficients than the previous theoretical results of \citet{mezei2019dissociative} when compared to recent measurements made at the Cryogenic Storage Ring \citep{paul2022experimental} over the astrophysically relevant temperature range for diffuse clouds ($\sim$40~K -- 130~K \citep{shull2021far}).

\section{On the Coulomb-Born Approximation for Rotational and vibrational (De-)Excitation}
\label{sec:coulomb}

The dipole moment of \ce{CH+} --- present in the molecular center-of-mass about which the molecule rotates --- couples partial waves of $\Delta l = \pm 1$, which reduces the accuracy of our partial wave basis ($l$=0--2) for such a long-range process as rotational excitation.
We include the effect of higher partial waves in the Coulomb-Born approximation \citep{boikova1968rotational,gailitis1976new,chu1974rotational}, similar to the method described in the work of \citet{rabadan1998rotions}, by calculating three different cross sections: cross sections obtained from our R-matrix method ($\sigma^\text{R-mat}$, calculated according to \ref{eqn:XS_rot}), total cross sections obtained in the Coulomb-Born approximation representing the contribution of all partial waves ($\sigma^\text{TCB}$, \ref{eqn:TCB}), and partial cross sections obtained in the Coulomb-Born approximation representing the contribution of the partial waves included in our basis ($\sigma^\text{PCB}$, \ref{eqn:PCB}).
The final rovibrational excitation cross sections are then a sum of the R-matrix cross sections and the difference between the total and partial Coulomb-Born cross sections, i.e.,
\begin{equation}
    \sigma^\text{RVE} = \sigma^\text{R-mat} + \sigma^\text{TCB} - \sigma^\text{PCB},
    \label{eqn:CB_Correction}.
\end{equation}

Lower partial-wave scattering is typically not well described by the Coulomb-Born approximation because the electron is too close to the molecule for the dipole interaction to be considered a perturbation.
This is especially true for $s$-wave scattering.
Therefore, we replace the $l=0$ -- 2 partial wave contribution from the Coulomb-Born approximation with those in our R-matrix calculations.
The partial Coulomb-Born cross sections are given by
\begin{equation}
    \begin{aligned}
        \sigma^\text{PCB}_{j'v'\leftarrow jv} &= 16 \pi \frac{k'}{k} \abs{\braket{v'|Q_\xi(R)|v}}^2
        \frac{2j'+1}{2\xi+1} {\small \pmat{j&j'&\xi\\0&0&0}} ^2
        \\
        &\times (2j+1)(2j'+1) \sum\limits_{ll'}^{l_\text{max}} {\small \pmat{l&l'&\xi\\0&0&0}}^2\abs{M_{ll'}^\xi}^2,
    \end{aligned}
    \label{eqn:PCB}
\end{equation}
where $l_\text{max}=2$ because our R-matrix calculations only include up to $l=2$ partial waves.
The dipole moment function is given by $Q_\xi(R)$ and the matrix elements $M^\xi_{ll'}$ are given by
\begin{equation}
    M^{\xi}_{ll'} = \frac{1}{kk'} \int\limits_0^\infty dr F_l(\eta,r) r^{-\xi-1} F_{l'}(\eta',r),
\end{equation}
where $F_{l}(\eta,r)$ is the regular radial Coulomb function, $\eta=-1/k$, and $\eta'=-1/k'$.

For an approach that does not treat vibration, the integral $\braket{v'|Q_\xi(R)|v}$ in (\ref{eqn:PCB}) can be replaced with the dipole moment at the equilibrium geometry of the ion.
Considering the dipolar coupling ($\xi=1$), the partial Coulomb-Born cross sections (\ref{eqn:PCB}) converge to the following as $l_\text{max} \to \infty$:
\begin{equation}
    \sigma^\text{TCB}_{j'v'\leftarrow jv} =  \frac{8}{3}\frac{\pi^3}{k^2} \abs{\braket{v'|Q_\xi(R)|v}}^2
    (2j'+1) {\small \pmat{j&j'&1\\0&0&0}}^2 f(\eta,\eta'),
    \label{eqn:TCB}
\end{equation}
where
\begin{equation}
    \begin{gathered}
        f(\eta,\eta') = \frac{e^{2\pi\eta}}{(e^{2\pi\eta}-1)(e^{2\pi\eta'}-1)} \chi_0 \frac{d}{d\chi_0} \abs{{}_2F_1(-i\eta,-i\eta';1;\chi_0)}^2,
        \\
        \zeta = \eta' - \eta, \quad
        \chi_0 = -4 \eta \eta' / \zeta^2.
    \end{gathered}
    \label{eqn:CB_f}
\end{equation}
The function ${}_2F_1(a,b;c;z)$ is the Gaussian hypergeometric function defined as:
\begin{equation}
    {}_2F_1(a,b;c;z) = \sum\limits_{n=0}^\infty \frac{(a)_n(b)_n}{(c)_n},
    \quad
    (q)_n = \frac{\Gamma(q+n)}{\Gamma(q)}.
\end{equation}

It should be noted that we only include the Coulomb-Born correction to $\Delta j=\pm1$ transitions; our $\Delta j=\pm2$ transitions are obtained purely from our R-matrix method.
For further detail in computing (\ref{eqn:CB_f}), we invite the reader to read the work of \citet{chu1974rotational}.
Additionally, the work of \citet{rabadan1998rotions} contains minor errors (non-squared Wigner 3-$j$ symbols) in their equations (3) and (4), which are corrected here.

\section{Rate Coefficients for Rovibronic (De-)Excitation}
\label{sec:RVERates}
\begin{figure}
  \centering
  \includegraphics[width=.48\textwidth]{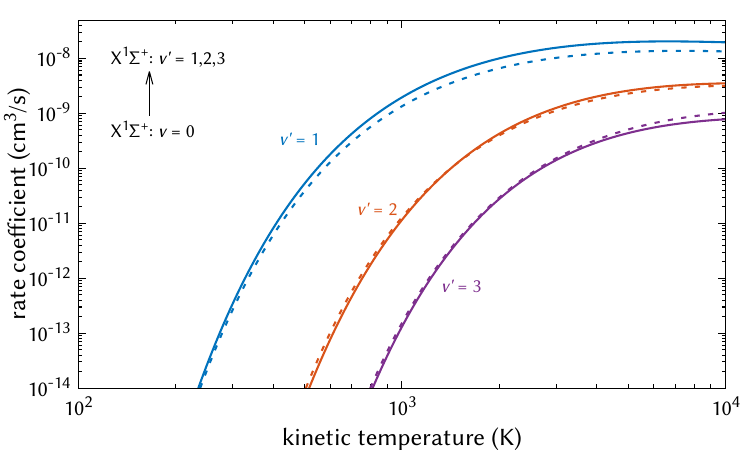}
  \caption{
    State-selected kinetic VE rate coefficients within the ground electronic state of \ce{CH+}.
    Solid lines represent rate coefficients from the present calculations, dashed lines are taken from a previous calculation \protect\citep{jiang2019cross}.
    The cross sections are obtained according to (\ref{eqn:XS_norot}) and the kinetic rates are obtained according to (\ref{eqn:kinetic_rate_select}).
  }
  \label{fig:kinetic_VE}
\end{figure}

\begin{figure}
  \centering
  \includegraphics[width=.48\textwidth]{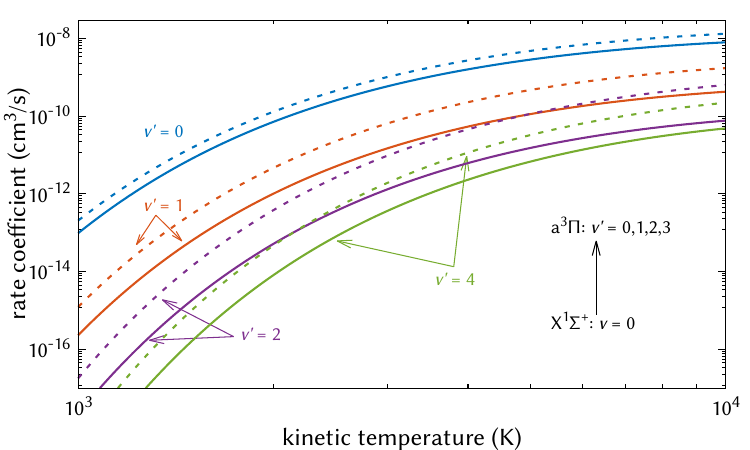}
  \caption{
    State-selected kinetic VE rate coefficients from the ground electronic state of \ce{CH+} to the first excited state of \ce{CH+}.
    Solid lines represent rate coefficients from the present calculations, dashed lines are taken from a previous calculation \protect\citep{jiang2019cross}.
    The cross sections are obtained according to (\ref{eqn:XS_norot}) and the kinetic rates are obtained according to (\ref{eqn:kinetic_rate_select}).
  }
  \label{fig:kinetic_VE2}
\end{figure}

\begin{figure}
  \centering
  \includegraphics[width=.48\textwidth]{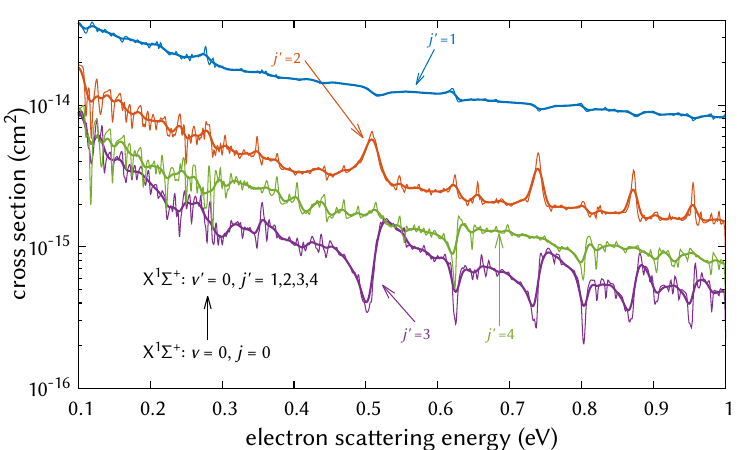}
  \caption{
    Rotational excitation cross sections within the ground vibronic state of \ce{CH+}.
    The cross sections are obtained according to (\ref{eqn:XS_rot}), and convolved with a Gaussian as per (\ref{eqn:convolution_gauss_numerical}) with $\gamma=1$~meV (thin lines) and $\gamma=5$meV (thick lines).
  }
  \label{fig:RE}
\end{figure}

\begin{figure}
  \centering
  \includegraphics[width=.48\textwidth]{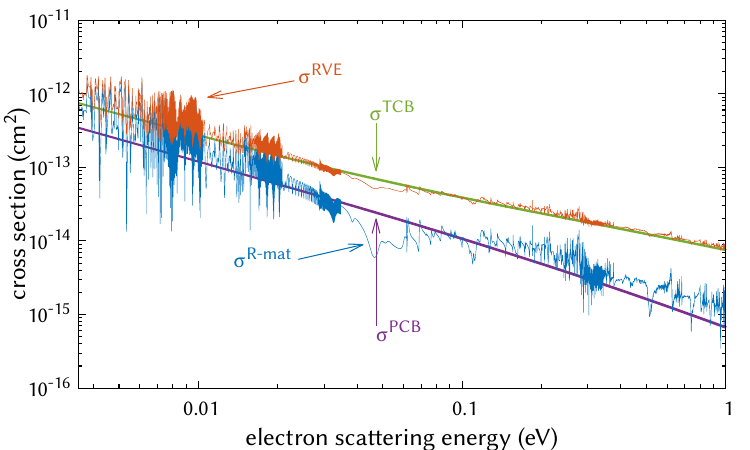}
  \caption{
    Comparison of cross sections for rotational excitation $j=0 \to j'=1$ obtained using the R-matrix approach with $s$, $p$, and $d$ partial waves ($\sigma^\text{R-mat}$) and the closed-form total Coulomb-Born approximation ($\sigma^{TCB}$). The figure shows also the partial Coulomb-Born cross section obtained with $s$, $p$, and $d$ partial waves ($\sigma^\text{PCB}$) and the cross section where the R matrix data is combined with the total Coulomb-Born cross section accounting for partial wave with $l>2$ ($\sigma^\text{RVE}$).
  }
  \label{fig:rot_exc_CB_corr}
\end{figure}

Fig. \ref{fig:kinetic_VE} shows the VE rate coefficients obtained by the present method.
The corresponding state-selected kinetic rate coefficients agree well with those of \citet{jiang2019cross}, as shown in Fig.~\ref{fig:kinetic_VE}.
In Fig. \ref{fig:kinetic_VE2}, we compare the present VE rate coefficients to the same work, starting from the ground vibronic state to the first excited electronic state.
The agreement is worse for excitation between electronic states, possibly due to our improved treatment of channels attached to the excited electronic states.
Fig. \ref{fig:RE} shows the present RE cross sections convolved with a Gaussian distribution according to (\ref{eqn:convolution_gauss_analytic},\ref{eqn:convolution_gauss_numerical}) with widths ($\gamma$) of 1~meV and 5~meV to demonstrate the smoothing out of resonances and to compare their overall magnitude.
The $\Delta j=\pm1$ transition is the largest over all displayed electron energies ($<1$~eV), as expected for such a strongly dipolar system as \ce{CH+}.
Fig. \ref{fig:rot_exc_CB_corr} shows the individual cross sections of (\ref{eqn:CB_Correction}), which show an increasing correction as a function of the incident electron energy.
The $\sigma^\text{RVE}$ and $\sigma^\text{R-mat}$ cross sections are not convolved, unlike Fig. \ref{fig:RE}.

\begin{figure}
  \centering
  \includegraphics[width=.48\textwidth]{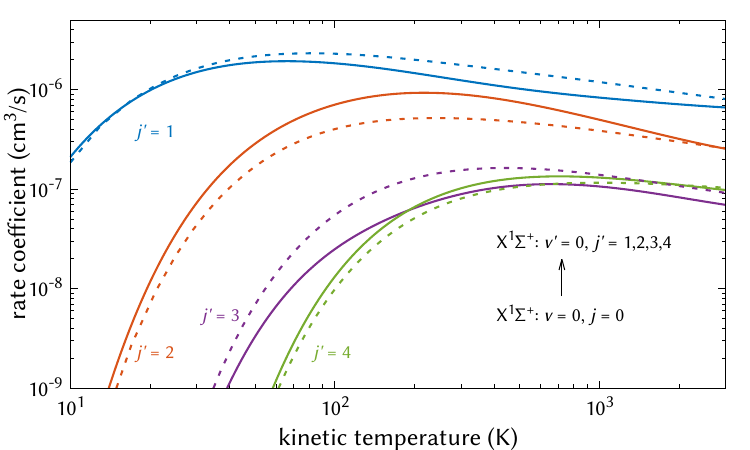}
  \caption{
    State-selected kinetic RE rate coefficients within the ground vibronic state of \ce{CH+}.
    Solid lines represent the present theory, dashed lines represent the results of \protect\citet{hamilton2016electron}.
  }
  \label{fig:kinetic_RE_compare}
\end{figure}

\begin{figure}
    \centering
    \includegraphics[width=0.48\textwidth]{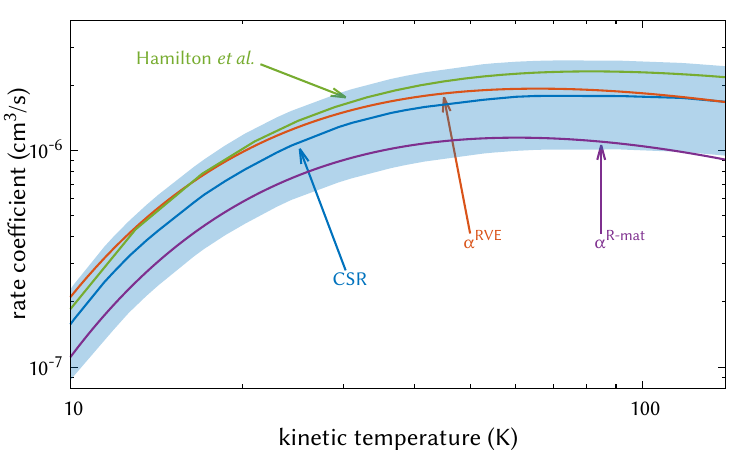}
    \caption{
      State-selected kinetic RE rate coefficients from $j=0$ to $j'=1$ within the ground vibronic state of \ce{CH+}.
      The measured rates coefficients from the CSR \protect\citep{kalosi2022laser} are compared to the results of \citet{hamilton2016electron} and our kinetic rate coefficients with ($\alpha^\text{RVE}$) and without ($\alpha^\text{R-mat}$) the Coulomb-Born correction.
    }
    \label{fig:kinetic_RE_compare_CSR}
\end{figure}

Fig. \ref{fig:kinetic_RE_compare} compares the present rotational excitation rate coefficients with those obtained by \citet{hamilton2016electron}, who included the Coulomb-Born correction for $\Delta j=\pm1,\pm2$ transitions.
We only include this correction for $\Delta j=\pm1$ transitions, but the agreement between the results is good overall.
\citet{hamilton2016electron} also use an R-matrix approach, but use the adiabatic-nuclei-rotation approximation to obtain rotational excitation (RE) cross sections and rate coefficients, while we use a frame transformation (\ref{eqn:RFT}) to describe the rotational structure of the ion.
Fig. \ref{fig:kinetic_RE_compare_CSR} illustrates the difference between the state-selected RE rates with and without the Coulomb-Born approximation, compared to the experimentally determined RE rate coefficient at the CSR \citep{kalosi2022laser} and again to the results of \citet{hamilton2016electron}.
Our present results with the Coulomb-Born correction show the best overall agreement with the CSR measurements, although all theoretical rates are within the provided 1-$\sigma$ uncertainty for most of the kinetic temperatures shown (10--140~K).


\section{Conclusions}
\label{sec:conclusion}

This paper presents kinetic state-selected DR, VE, and RE rate coefficients obtained with our DR method \citep{forer2023unified}, which also allows us to calculate RVE cross sections and rate coefficients with little extra effort.
The DR rate coefficients agree well overall with recent experimental measurements made at the Cryogenic Storage Ring \citep{paul2022experimental} and better than previous theoretical treatments.
Our VE rate coefficients, compared to the work of \citet{jiang2019cross}, agree well for vibrational excitation within the ground electronic state of \ce{CH+}.
However, the results differ by up to an order of magnitude for vibronic excitation to the first excited state of \ce{CH+}, which we attribute to a more accurate description of channels attached to excited electronic states \citep{forer2023unified}.

We also obtain RE rate coefficients within the ground electronic state of \ce{CH+}, which we compare to the work of \citet{hamilton2016electron} which is an R-matrix method that describes rotational excitation with the adiabatic-nuclei-rotation approximation and does not treat vibration.
They correct their $\Delta j=\pm1,\pm2$ transitions with the Coulomb-Born approximation, while we only do so for $\Delta j=\pm1$ transitions.
Results between our approaches agree well over the presented kinetic temperatures.
Compared to $j=0\to j'=1$ rate coefficients recently measured at the CSR \citep{kalosi2022laser}, our theoretical results using the Coulomb-Born correction agree better over all plotted kinetic temperatures than our theoretical results without the Coulomb-Born correction, and slightly better than the recent theoretically determined rate coefficients of \citet{hamilton2016electron} over most kinetic temperatures between 10~K and 140~K.
However, all theoretically determined rates under 100~K are within the experimental uncertainty.

\section*{Acknowledgements}
We are thankful for the support from the National Science Foundation, Grant Nos.2110279 (UCF) and 2102187 (Purdue), the Fulbright-University of Bordeaux Doctoral Research Award, and the program \quotes{Accueil des chercheurs étrangers} of CentraleSupélec.

\section*{Data Availability}
State-selected kinetic rate coefficients from the present calculations are included in the supplementary materials.

\bibliographystyle{unsrtnat}
\bibliography{refs.bib}

\providecommand{\noopsort}[1]{}\providecommand{\singleletter}[1]{#1}%
\begin{thebibliography}{27}
\providecommand{\natexlab}[1]{#1}
\providecommand{\url}[1]{\texttt{#1}}
\expandafter\ifx\csname urlstyle\endcsname\relax
  \providecommand{\doi}[1]{doi: #1}\else
  \providecommand{\doi}{doi: \begingroup \urlstyle{rm}\Url}\fi

\bibitem[Douglas and Herzberg(1941)]{douglas1941note}
AE~Douglas and G~Herzberg.
\newblock Note on {CH}$^+$ in interstellar space and in the laboratory.
\newblock \emph{Astrophys. J.}, 94:\penalty0 381, 1941.

\bibitem[Adams(1941)]{adams1941some}
Walter~S Adams.
\newblock Some results with the {COUD{\'E}} spectrograph of the mount wilson
  observatory.
\newblock \emph{Astrophys. J.}, 93:\penalty0 11, 1941.

\bibitem[Hierl et~al.(1997)Hierl, Morris, and Viggiano]{hierl1997rate}
Peter~M Hierl, Robert~A Morris, and AA~Viggiano.
\newblock Rate coefficients for the endothermic reactions {C$^+$($^2$P)+ H$_2$
  (D$_2$)$\to$CH$^+$(CD$^+$)+H (D)} as functions of temperature from
  400--1300~{K}.
\newblock \emph{J. Chem. Phys.}, 106\penalty0 (24):\penalty0 10145--10152,
  1997.

\bibitem[Shull et~al.(2021)Shull, Danforth, and Anderson]{shull2021far}
J~Michael Shull, Charles~W Danforth, and Katherine~L Anderson.
\newblock A far ultraviolet spectroscopic explorer survey of interstellar
  molecular hydrogen in the galactic disk.
\newblock \emph{Astrophys. J.}, 911\penalty0 (1):\penalty0 55, 2021.

\bibitem[Godard and Cernicharo(2013)]{godard2013complete}
Benjamin Godard and Jos{\'e} Cernicharo.
\newblock A complete model of {CH}$^+$ rotational excitation including
  radiative and chemical pumping processes.
\newblock \emph{Astron. Astrophys.}, 550:\penalty0 A8, 2013.

\bibitem[Amitay et~al.(1996)Amitay, Zajfman, Forck, Hechtfischer, Seidel,
  Grieser, Habs, Repnow, Schwalm, and Wolf]{amitay1996dissociative}
Z~Amitay, D~Zajfman, P~Forck, U~Hechtfischer, B~Seidel, M~Grieser, D~Habs,
  R~Repnow, D~Schwalm, and A~Wolf.
\newblock Dissociative recombination of {CH$^+$}: Cross section and final
  states.
\newblock \emph{Phys. Rev. A}, 54\penalty0 (5):\penalty0 4032, 1996.

\bibitem[Paul et~al.(2022)Paul, Grieser, Grussie, von Hahn, Isberner, Ábel
  Kálosi, Krantz, Kreckel, Müll, Neufeld, Savin, Schippers, Wilhelm, Wolf,
  Wolfire, and Novotný]{paul2022experimental}
Daniel Paul, Manfred Grieser, Florian Grussie, Robert von Hahn, Leonard~W.
  Isberner, Ábel Kálosi, Claude Krantz, Holger Kreckel, Damian Müll,
  David~A. Neufeld, Daniel~W. Savin, Stefan Schippers, Patrick Wilhelm, Andreas
  Wolf, Mark~G. Wolfire, and Oldřich Novotný.
\newblock Experimental determination of the dissociative recombination rate
  coefficient for rotationally cold {CH}$^+$ and its implications for diffuse
  cloud chemistry.
\newblock \emph{Astrophys. J.}, 939\penalty0 (2):\penalty0 122, nov 2022.
\newblock \doi{10.3847/1538-4357/ac8e02}.
\newblock URL \url{https://dx.doi.org/10.3847/1538-4357/ac8e02}.

\bibitem[K{\'a}losi et~al.(2022)K{\'a}losi, Grieser, von Hahn, Hechtfischer,
  Krantz, Kreckel, M{\"u}ll, Paul, Savin, Wilhelm, Wolf, Wolfire, and
  Novotn\'y]{kalosi2022laser}
{\'A}bel K{\'a}losi, Manfred Grieser, Robert von Hahn, Ulrich Hechtfischer,
  Claude Krantz, Holger Kreckel, Damian M{\"u}ll, Daniel Paul, Daniel~W Savin,
  Patrick Wilhelm, Andreas Wolf, Mark~G. Wolfire, and Oldrich Novotn\'y.
\newblock Laser probing of the rotational cooling of molecular ions by electron
  collisions.
\newblock \emph{Phys. Rev. Lett.}, 128\penalty0 (18):\penalty0 183402, 2022.

\bibitem[Forer et~al.(2023)Forer, Hvizdoš, Jiang, Ayouz, Greene, and
  Kokoouline]{forer2023unified}
Joshua Forer, Dávid Hvizdoš, Xianwu Jiang, Mehdi Ayouz, Chris~H. Greene, and
  Viatcheslav Kokoouline.
\newblock Unified treatment of resonant and non-resonant mechanisms in
  dissociative recombination: benchmark study of {CH}$^+$.
\newblock \emph{Phys. Rev. A}, 85:\penalty0 5430, 2023.

\bibitem[Hamilton and Greene(2002)]{hamilton2002competition}
Edward~L Hamilton and Chris~H Greene.
\newblock Competition among molecular fragmentation channels described with
  siegert channel pseudostates.
\newblock \emph{Phys. Rev. Lett.}, 89\penalty0 (26):\penalty0 263003, 2002.

\bibitem[Kokoouline and Greene(2003)]{kokoouline2003unified}
Viatcheslav Kokoouline and Chris~H Greene.
\newblock Unified theoretical treatment of dissociative recombination of
  {$D_{3h}$} triatomic ions: Application to {H$_3^+$} and {D$_3^+$}.
\newblock \emph{Phys. Rev. A}, 68\penalty0 (1):\penalty0 012703, 2003.

\bibitem[{\v{C}}ur{\'\i}k et~al.(2020){\v{C}}ur{\'\i}k, Hvizdo{\v{s}}, and
  Greene]{vcurik2020dissociative}
Roman {\v{C}}ur{\'\i}k, D{\'a}vid Hvizdo{\v{s}}, and Chris~H Greene.
\newblock Dissociative recombination of cold {HeH}$^+$ ions.
\newblock \emph{Phys. Rev. Lett.}, 124\penalty0 (4):\penalty0 043401, 2020.

\bibitem[Chu and Dalgarno(1974)]{chu1974rotational}
Shih-I Chu and A.~Dalgarno.
\newblock Rotational excitation of {CH}$^+$ by electron impact.
\newblock \emph{Phys. Rev. A}, 10:\penalty0 788--792, Sep 1974.
\newblock \doi{10.1103/PhysRevA.10.788}.
\newblock URL \url{https://link.aps.org/doi/10.1103/PhysRevA.10.788}.

\bibitem[Dickinson and Munoz(1977)]{dickinson1977rotational}
AS~Dickinson and JM~Munoz.
\newblock Rotational excitation of polar molecular ions by slow electrons.
\newblock \emph{J. Phys. B: At. Mol. Phys.}, 10\penalty0 (15):\penalty0 3151,
  1977.

\bibitem[Lim et~al.(1999)Lim, Rabad{\'a}n, and Tennyson]{lim1999electron}
Andrew~J Lim, Ismanuel Rabad{\'a}n, and Jonathan Tennyson.
\newblock Electron-impact rotational excitation of {CH}$^+$.
\newblock \emph{Mon. Not. R. Astron. Soc.}, 306\penalty0 (2):\penalty0
  473--478, 1999.

\bibitem[Flower(1979)]{flower1979electron}
D~R Flower.
\newblock Electron collisional excitation of rotational transitions in {CH}$^+$
  and {HeH}$^+$.
\newblock \emph{Astron. Astrophys.}, 73:\penalty0 237--239, 1979.

\bibitem[Jiang et~al.(2019)Jiang, Yuen, Cortona, Ayouz, and
  Kokoouline]{jiang2019cross}
Xianwu Jiang, Chi~Hong Yuen, Pietro Cortona, Mehdi Ayouz, and Viatcheslav
  Kokoouline.
\newblock Cross sections for vibronic excitation of {CH}$^+$ by low-energy
  electron impact.
\newblock \emph{Phys. Rev. A}, 100\penalty0 (6):\penalty0 062711, 2019.

\bibitem[Carr et~al.(2012)Carr, Galiatsatos, Gorfinkiel, Harvey, Lysaght,
  Madden, Ma{\v{s}}{\'\i}n, Plummer, Tennyson, and Varambhia]{carr2012ukrmol}
JM~Carr, PG~Galiatsatos, Jimena~D Gorfinkiel, Alex~G Harvey, MA~Lysaght,
  D~Madden, Z~Ma{\v{s}}{\'\i}n, M~Plummer, Jonathan Tennyson, and HN~Varambhia.
\newblock {UKRmol}: a low-energy electron-and positron-molecule scattering
  suite.
\newblock \emph{Eur. Phys. J. D.}, 66\penalty0 (3):\penalty0 58, 2012.

\bibitem[Seaton(1983)]{seaton1983quantum}
MJ~Seaton.
\newblock Quantum defect theory.
\newblock \emph{Rep. Prog. Phys.}, 46\penalty0 (2):\penalty0 167, 1983.

\bibitem[Aymar et~al.(1996)Aymar, Greene, and
  Luc-Koenig]{aymar1996multichannel}
Mireille Aymar, Chris~H Greene, and Eliane Luc-Koenig.
\newblock Multichannel rydberg spectroscopy of complex atoms.
\newblock \emph{Rev. Mod. Phys.}, 68\penalty0 (4):\penalty0 1015, 1996.

\bibitem[Tennyson(2010)]{tennyson2010electron}
Jonathan Tennyson.
\newblock Electron--molecule collision calculations using the {R}-matrix
  method.
\newblock \emph{Phys. Rep.}, 491\penalty0 (2-3):\penalty0 29--76, 2010.

\bibitem[Tennyson et~al.(2007)Tennyson, Brown, Munro, Rozum, Varambhia, and
  Vinci]{tennyson2007quantemol}
Jonathan Tennyson, Daniel~B Brown, James~J Munro, Iryna Rozum, Hemal~N
  Varambhia, and Natalia Vinci.
\newblock Quantemol-{N}: an expert system for performing electron molecule
  collision calculations using the {R}-matrix method.
\newblock \emph{J. Phys. Conf. Ser.}, 86\penalty0 (1):\penalty0 012001, 2007.

\bibitem[Mezei et~al.(2019)Mezei, Ep{\'e}e~Ep{\'e}e, Motapon, and
  Schneider]{mezei2019dissociative}
Zsolt~J Mezei, Michel~D Ep{\'e}e~Ep{\'e}e, Ousmanou Motapon, and Ioan~F
  Schneider.
\newblock Dissociative recombination of {CH$^+$} molecular ion induced by very
  low energy electrons.
\newblock \emph{Atoms}, 7\penalty0 (3):\penalty0 82, 2019.

\bibitem[Boikova and Ob'edkov(1968)]{boikova1968rotational}
RF~Boikova and VD~Ob'edkov.
\newblock Rotational and vibrational excitation of molecular ions by electrons.
\newblock \emph{Soviet Phys. JETP}, 54:\penalty0 1439, 1968.

\bibitem[Gailitis(1976)]{gailitis1976new}
M~Gailitis.
\newblock New forms of asymptotic expansions for wavefunctions of
  charged-particle scattering.
\newblock \emph{J. Phys. B}, 9\penalty0 (5):\penalty0 843, 1976.

\bibitem[Rabad{\'a}n and Tennyson(1998)]{rabadan1998rotions}
Ismanuel Rabad{\'a}n and Jonathan Tennyson.
\newblock Rotions: A program for the calculation of rotational excitation cross
  sections in electron—molecular ion collisions.
\newblock \emph{Comput. Phys. Commun.}, 114\penalty0 (1-3):\penalty0 129--141,
  1998.

\bibitem[Hamilton et~al.(2015)Hamilton, Faure, and
  Tennyson]{hamilton2016electron}
James~R. Hamilton, Alexandre Faure, and Jonathan Tennyson.
\newblock {Electron-impact excitation of diatomic hydride cations – I.
  {HeH}$^+$, {CH}$^+$, {ArH}$^+$}.
\newblock \emph{Mon. Not. R. Astron. Soc.}, 455\penalty0 (3):\penalty0
  3281--3287, 11 2015.
\newblock ISSN 0035-8711.
\newblock \doi{10.1093/mnras/stv2429}.

\end{thebibliography}

\section*{Supplementary Material}
Supplementary Material is available in the online version of this article

\end{document}